\documentclass[11pt,reqno]{amsart}
\usepackage{amssymb}
\usepackage{amsfonts}
\usepackage{amsmath}
\usepackage{stmaryrd}
\usepackage{physics}
\usepackage{braket}
\usepackage{dsfont}
\usepackage{bbold}
\usepackage{graphicx}
\usepackage{relsize}
\usepackage[table,xcdraw]{xcolor}
\usepackage{enumerate}
\usepackage{hyperref}
\hypersetup{
    colorlinks=true,
    linkcolor=blue,
    citecolor=red,
    filecolor=magenta,      
    urlcolor=black,
}
\usepackage[margin=1in,top=0.8in,bottom=0.8in]{geometry}
\usepackage{enumitem}
\usepackage{mathtools}
\usepackage{graphbox}
\usepackage{comment}
\usepackage{float}
\usepackage[font=scriptsize]{caption}

\renewcommand{\epsilon}{\varepsilon}
\renewcommand{\phi}{\varphi}

\newcommand{\M}[1]{\mathcal{M}_{#1}}
\newcommand{\C}[1]{\mathbb{C}^{#1}}
\newcommand{\N}{\mathbb{N}}

\DeclareMathOperator{\ident}{id}
\DeclareMathOperator{\Tra}{Tr}
\newcommand{\ra}{\rightarrow}

\newtheorem{theorem}{Theorem}[]
\newtheorem{definition}[theorem]{Definition}
\newtheorem*{definition*}{Definition}

\newtheorem{corollary}[theorem]{Corollary}
\newtheorem{lemma}[theorem]{Lemma}

\newtheorem{question}[theorem]{Question}

\newtheorem*{conjecture*}{Conjecture}

\theoremstyle{definition}

\definecolor{darkgreen}{rgb}{0,0.392,0}

\title{Bi-PPT channels are entanglement breaking}

\begin{document}
\begin{abstract}
In~\cite{hirche2022bounding}, Hirche and Leditzky introduced the notion of bi-PPT channels which are quantum channels that stay completely positive under composition with a transposition and such that the same property holds for one of their complementary channels. They asked whether there are examples of such channels that are not antidegradable. We show that this is not the case, since bi-PPT channels are always entanglement breaking. We also show that degradable quantum channels staying completely positive under composition with a transposition are entanglement breaking.
\end{abstract}

\footnotetext[1]{Corresponding author.}

\author{Alexander Müller-Hermes$^2$}
\address{\small{\parbox{\linewidth}{Department of Mathematics, \\ University of Oslo, P.O. box 1053, Blindern, 0316 Oslo, Norway. \vspace{0.1cm}}}}
\email{muellerh@math.uio.no}

\author{Satvik Singh$^{1,2}$}
\address{\small{\parbox{\linewidth}{Department of Applied Mathematics and Theoretical Physics, \\ University of Cambridge, Cambridge, United Kingdom. \vspace{0.1cm}}}}
\email{satviksingh2@gmail.com}
\maketitle

\vspace{-0.2cm}

We denote by $\M{d}$ the set of $d\times d$ matrices with complex entries and by $\M{d}^+\subset \M{d}$ the cone of positive semidefinite matrices. A linear map $\Phi:\M{d_A}\ra \M{d_B}$ is called \emph{positive} if $\Phi(\M{d_A}^+)\subseteq \M{d_B}^+$ and \emph{completely positive} (CP) if $\ident_n\otimes \,\Phi$ is a positive map for every $n\in\N$, where $\ident_n$ denotes the identity map on $\M{n}$. Moreover, $\Phi$ is called \emph{completely copositive} (coCP) if $\vartheta_n\otimes \Phi$ is positive for every $n\in\N$, where $\vartheta_n$ denotes the matrix transpose on $\M{n}$ in some fixed basis. We will call $\Phi$ \emph{PPT} if it is both CP and coCP. Finally, $\Phi$ is called \emph{entanglement breaking} if for all $X\in (\M{n}\otimes \M{d_A})^+$ and $n\in \N$, $(\ident_n\otimes \,\Phi) (X)$ is \emph{separable}, i.e., it lies in the convex hull of $Y\otimes Z$, $Y\in \M{n}^+, Z\in \M{d_B}^+$. The \emph{Choi matrix} of a linear map $\Phi:\M{d_A}\to \M{d_B}$ is defined as
\begin{equation}
    J(\Phi) := \sum_{i,j=1}^{d_A} \ketbra{i}{j} \otimes \Phi (\ketbra{i}{j}) \in \M{d_A}\otimes \M{d_B}.
\end{equation}
It is known that a linear map $\Phi$ is PPT if and only if its Choi matrix $J(\Phi)$ is positive and has positive partial transpose (PPT). Similarly, $\Phi$ is entanglement breaking if and only if $J(\Phi)$ is separable. When studying the quantum capacity of quantum channels (i.e. CP trace preserving linear maps), PPT channels appear naturally as one of the classes of zero-capacity channels \cite{Smith2012incapacity}. Another notion motivated by the capacity problem is that of a complementary pair of quantum channels~\cite{devetak2005capacity,holevo2007complementary,King2007complement}. We will use a slightly more general form of this notion below.

\begin{definition}
Two CP linear maps $\Phi:\M{d_A}\to \M{d_B}$ and $\Psi:\M{d_A}\to \M{d_C}$ are said to be \emph{complementary} to each other if there exists a linear operator $L:\C{d_A}\to \C{d_B}\otimes \C{d_C}$ such that
\begin{equation*}
    \forall X\in \M{d_A}: \quad \Phi(X) = \Tra_C (LXL^\dagger) \quad\text{and}\quad \Psi(X) = \Tra_B (LXL^\dagger),
\end{equation*}
where $\Tra_B$ and $\Tra_C$ denote the partial traces over the subsystems with the indicated labels.
\end{definition}

Let ($\Phi$, $\Psi$) be a complementary pair of CP maps. We call $\Phi$ \emph{antidegradable}~\cite{cubitt2008structure} (and $\Psi$ \emph{degradable}~\cite{devetak2005capacity}) if there exists a CP map $\Omega$ such that such that $\Phi = \Omega \circ \Psi$. Antidegradable quantum channels form the second known class of channels with no quantum capacity. The classes of PPT and antidegradable channels are incomparable in general, since there are known examples of channels that lie in either one of the classes but not the other. Entanglement breaking channels are known to be both PPT and antidegradable. The following question was asked in~\cite{hirche2022bounding}:

\begin{question}
Are there any pairs of complementary quantum channels $\Phi$ and $\Psi$ such that both $\Phi$ and $\Psi$ are PPT, but that at least one of them is not antidegradable?
\end{question}

Hirche and Leditzky~\cite{hirche2022bounding} argue that such examples of channels could be used to exhibit superactivation of the private capacity of quantum channels, since they would have vanishing private capacity. Unfortunately, we can prove that no such example exists. Our proof exploits the following two facts about the entanglement properties of low rank quantum states. Recall that a positive matrix $X\in (\M{d_A}\otimes \M{d_B})^+$ is called \emph{undistillable} if no pure entanglement can be asymptotically extracted from any number of copies of $X$ by local operations and 2-way classical communication (2-LOCC) \cite{Horodecki1998distillation}. 

\begin{lemma} \label{lemma:1} \cite{Horodecki2003entanglement}
Let $X\in (\M{d_A}\otimes \M{d_B})^+$ be a positive matrix such that 
\begin{equation*}
    \operatorname{rank}X < \max\{\operatorname{rank} \Tra_A X, \operatorname{rank} \Tra_B X \}.
\end{equation*}
Then, $X$ is distillable. In particular, $X$ cannot be PPT whenever this is the case.
\end{lemma}

\begin{lemma}\label{lemma:2}
Let $X\in (\M{d_A}\otimes \M{d_B})^+$ be a positive matrix such that 
\begin{equation*}
    \operatorname{rank}X \leq \max\{\operatorname{rank}\Tra_A X, \operatorname{rank} \Tra_B X \}.
\end{equation*}
Then, $X$ is separable if and only if it is PPT if and only if it is undistillable.
\end{lemma}
\begin{proof}
The equivalence of PPT and separability is established in \cite{Horodecki2000entanglement}. Moreover, if $X$ is PPT, then it is clearly undistillable. Conversely, if $X$ is undistillable, then Lemma~\ref{lemma:1} shows that $\operatorname{rank}X = \max\{\operatorname{rank}\Tra_A X, \operatorname{rank} \Tra_B X \}$ and \cite[Theorem 10]{Chen2011entanglement} implies that $X$ must be PPT.
\end{proof}

We can now state and prove our main results.
\begin{theorem}\label{theorem:main}
Let $\Phi:\M{d_A}\to \M{d_B}$ be a PPT linear map. Then, for any CP linear map $\Psi:\M{d_A}\to \M{d_C}$ that is complementary to $\Phi$, the following are equivalent:
\begin{itemize}
    \item $\Psi$ is PPT.
    \item $\Psi$ is entanglement breaking.
    \item $J(\Psi)$ is undistillable.
\end{itemize}
Moreover, if $\Phi$ is PPT but not entanglement breaking, then $J(\Psi)$ is distillable \footnote{After this paper was posted online, we were made aware of an earlier result \cite[Theorem 2]{Chen2011entanglement2} which establishes the same equivalences as in Theorem~\ref{theorem:main} (with identical assumption on $\Phi$) but is formulated in the language of quantum states. More recently, the results of \cite{Singh2022main} show that the conclusion of Theorem~\ref{theorem:main} remains valid under a much weaker assumption on $\Phi$, namely that $\Phi$ has zero (1-way) quantum capacity.}.
\end{theorem}
\begin{proof}
By definition, $J(\Phi)$ and $J(\Psi)$ admit a common purification $\ket{L}\in \C{d_A}\otimes\C{d_B}\otimes \C{d_C}$, i.e., $J(\Phi) = L_{AB} = \Tra_C \ketbra{L}$ and $J(\Psi) = L_{AC} = \Tra_B \ketbra{L}$. Now, since $\Phi$ is PPT, Lemma~\ref{lemma:1} implies that $\operatorname{rank} L_{AB} \geq  \operatorname{rank} L_B$, where $L_B=\Tra_A (L_{AB})$. Hence, by exploiting some elementary properties of purifications, we see that
\begin{equation}
    \operatorname{rank} L_C =\operatorname{rank} L_{AB} \geq  \operatorname{rank} L_B = \operatorname{rank} L_{AC},
\end{equation}
where $L_C=\Tra_A (L_{AC})$. Applying Lemma~\ref{lemma:2} on $J(\Psi)=L_{AC}$ then establishes the desired equivalences. Finally, if $\Phi$ is not entanglement breaking, then Lemma~\ref{lemma:2} shows that the above rank inequality becomes strict, i.e.,
\begin{equation}
    \operatorname{rank} L_C =\operatorname{rank} L_{AB} >  \operatorname{rank} L_B = \operatorname{rank} L_{AC},
\end{equation}
in which case applying Lemma~\ref{lemma:1} on $J(\Psi)=L_{AC}$ shows that $J(\Psi)$ is distillable.
\end{proof}

\begin{corollary}\label{corollary:1}
If two PPT linear maps $\Phi:\M{d_A}\to \M{d_B}$ and $\Psi:\M{d_A}\to \M{d_C}$ are complementary to each other, then they must both be entanglement breaking.
\end{corollary}
\begin{proof}
The corollary trivially follows from Theorem~\ref{theorem:main}.
\end{proof}

A simple application of our results suffices to prove that all degradable PPT linear maps must be entanglement breaking. This generalizes the well-known fact that Schur multiplier maps are entanglement breaking whenever they are PPT (see, e.g.,~\cite{kennedy2018composition} or \cite[Example 7.3]{Singh2021diagonalunitary}).

\begin{corollary}\label{corollary:2}
Let ($\Phi$, $\Psi$) be a complementary pair of CP linear maps such that $\Phi=\Omega\circ \Psi$ for some CP linear map $\Omega$. Then, if $\Psi$ is PPT, both $\Phi$ and $\Psi$ must be entanglement breaking.  
\end{corollary}
\begin{proof}
Clearly, since $\Psi$ is assumed to be PPT, $\Phi=\Omega\circ \Psi$ must be PPT as well. Corollary~\ref{corollary:1} then implies that both $\Phi$ and $\Psi$ are entanglement breaking.
\end{proof}

The above corollary can be used to construct some examples of pairs of complementary PPT linear maps $(\Phi, \Psi)$ such that both $\Phi$ and $\Psi$ are entanglement breaking. One simply needs to choose $\Psi$ to be PPT and degradable. A canonical choice is to define $\Psi:\M{d}\to \M{d}$ as
\begin{equation}
    \forall X\in \M{d}: \quad \Psi(X) = T\odot X,
\end{equation}
where $T\in \M{d}^+$ is a diagonal matrix with non-negative entries and $\odot$ denotes the entrywise matrix product. It is then easy to check that $\Psi$ is a PPT Schur multiplier map. Moreover, since Schur multiplier maps are degradable \cite{devetak2005capacity}, Corollary~\ref{corollary:2} informs us that both $\Psi$ and any CP linear map $\Phi$ that is complementary to $\Psi$ must be entanglement breaking.

\vspace{0.5cm}

\noindent\textbf{Acknowledgements.} We thank Felix Leditzky and Christoph Hirche for their helpful feedback on a first draft of this paper. We also thank the Centre International de Rencontres Mathématiques (CIRM) for their hospitality during the workshop ``Random tensors and related topics" where this work was initiated.

\vspace{0.25cm}

\noindent\textbf{Conflict of interest.} On behalf of all authors, the corresponding author states that there is no conflict of interest.

\vspace{0.25cm}

\noindent\textbf{Data availability.} No data sets were generated during this study.

\bibliographystyle{alpha}
\bibliography{references}

\begin{thebibliography}{KMNR07}

\bibitem[C{\DJ}11]{Chen2011entanglement}
Lin Chen and Dragomir~{\v{Z}} {\DJ}okovi{\'{c}}.
\newblock Distillability and {PPT} entanglement of low-rank quantum states.
\newblock {\em Journal of Physics A: Mathematical and Theoretical},
  44(28):285303, June 2011.

\bibitem[CRS08]{cubitt2008structure}
Toby~S Cubitt, Mary~Beth Ruskai, and Graeme Smith.
\newblock The structure of degradable quantum channels.
\newblock {\em Journal of Mathematical Physics}, 49(10):102104, 2008.

\bibitem[DS05]{devetak2005capacity}
Igor Devetak and Peter~W Shor.
\newblock The capacity of a quantum channel for simultaneous transmission of
  classical and quantum information.
\newblock {\em Communications in Mathematical Physics}, 256(2):287--303, 2005.

\bibitem[HC11]{Chen2011entanglement2}
Masahito Hayashi and Lin Chen.
\newblock Weaker entanglement between two parties guarantees stronger
  entanglement with a third party.
\newblock {\em Phys. Rev. A}, 84:012325, Jul 2011.

\bibitem[HHH98]{Horodecki1998distillation}
Micha\l{} Horodecki, Pawe\l{} Horodecki, and Ryszard Horodecki.
\newblock Mixed-state entanglement and distillation: Is there a ``bound''
  entanglement in nature?
\newblock {\em Phys. Rev. Lett.}, 80:5239--5242, Jun 1998.

\bibitem[HL22]{hirche2022bounding}
Christoph Hirche and Felix Leditzky.
\newblock Bounding quantum capacities via partial orders and complementarity.
\newblock {\em preprint arXiv:2202.11688}, 2022.

\bibitem[HLVC00]{Horodecki2000entanglement}
Pawe\l{} Horodecki, Maciej Lewenstein, Guifr\'e Vidal, and Ignacio Cirac.
\newblock Operational criterion and constructive checks for the separability of
  low-rank density matrices.
\newblock {\em Phys. Rev. A}, 62:032310, Aug 2000.

\bibitem[Hol07]{holevo2007complementary}
Alexander~S Holevo.
\newblock Complementary channels and the additivity problem.
\newblock {\em Theory of Probability \& Its Applications}, 51(1):92--100, 2007.

\bibitem[HSTT03]{Horodecki2003entanglement}
Pawel Horodecki, John~A Smolin, Barbara~M Terhal, and Ashish~V Thapliyal.
\newblock Rank two bipartite bound entangled states do not exist.
\newblock {\em Theoretical Computer Science}, 292(3):589--596, January 2003.

\bibitem[KMNR07]{King2007complement}
C.~King, K.~Matsumoto, M.~Nathanson, and M.B. Ruskai.
\newblock Properties of conjugate channels with applications to additivity and
  multiplicativity.
\newblock {\em Markov Processes And Related Fields}, 13(2):391--423, 2007.

\bibitem[KMP18]{kennedy2018composition}
Matthew Kennedy, Nicholas~A Manor, and Vern~I Paulsen.
\newblock {Composition of PPT maps}.
\newblock {\em Quantum Information \& Computation}, 18(5-6):472--480, 2018.

\bibitem[SD22]{Singh2022main}
Satvik Singh and Nilanjana Datta.
\newblock Fully undistillable quantum states are separable.
\newblock {\em preprint arxiv:2207:05193}, 2022.

\bibitem[SN21]{Singh2021diagonalunitary}
Satvik Singh and Ion Nechita.
\newblock Diagonal unitary and orthogonal symmetries in quantum theory.
\newblock {\em {Quantum}}, 5:519, August 2021.

\bibitem[SS12]{Smith2012incapacity}
Graeme Smith and John~A. Smolin.
\newblock Detecting incapacity of a quantum channel.
\newblock {\em Phys. Rev. Lett.}, 108:230507, Jun 2012.

\end{thebibliography}
\vspace{0.25cm}
\hrule

\end{document}